\documentstyle[prb,preprint,aps]{revtex}
\begin{document}
\draft
\title{\bf Structure of aluminum atomic chains}
\author{Prasenjit Sen$^{(1)}$, S. Ciraci$^{(1,2)}$, A. Buldum$^{(3)}$,
Inder P. Batra$^{(1)}$} 
\address{$^{(1)}$ Department of Physics, University of Illinois at Chicago,
Chicago IL 60607-7059}
\address{$^{(2)}$ Department of Physics, Bilkent University, Bilkent,
Ankara 06533, Turkey}
\address{$^{(3)}$ Department of Physics and Astronomy, The University of 
North Carolina at Chapel Hill, Chapel Hill, NC 27599}
\begin{titlepage}
\maketitle
\begin{abstract}
First-principles density functional calculations reveal that aluminum
can form planar chains in zigzag and ladder structures. The most 
stable one has equilateral triangular geometry with four 
nearest neighbors; the other stable zigzag structure has wide 
bond angle and allows for two nearest neighbors. An intermediary
structure has the ladder geometry and is formed by two strands. 
While all these planar geometries are more favored energetically  
than the linear chain, the binding becomes even stronger in non-planar
geometries. We found that by going from bulk to a chain
the character of bonding changes and acquires directionality. The 
conductance of zigzag and linear chains is $4e^2/h$ under ideal
ballistic conditions.
\end{abstract} 
\end{titlepage}

\section{introduction}

The fabrication of the stable gold monoatomic chains suspended between two 
gold electrodes is one of the milestones in 
nanoscience.\cite{ohnish,yanson} Issues brought about by this achievement
are yet to be resolved: Stable chains were obtained by 
stretching gold nanowires; no other metal, such as Al, Cu, has been observed
to form a stable monoatomic chain yet. The monoatomic chain, being an 
ultimate one-dimensional (1D) structure, has been a testing ground 
for the theories and concepts developed earlier for three-dimensional (3D)  
systems. For example, it is of fundamental importance to know 
the atomic structure in a truly 1D nanowire and how the mechanical 
and electronic properties change in the lower dimensionality.

The density 
functional theory has been successful in predicting electronic 
and mechanical properties of bulk metals, where each atom has 8-12 
nearest neighbors depending on the crystal structure. While 
many neighbors in a 3D structure is a signature of the formation of
metallic bonds, it is not obvious whether the ``metallic'' bond picture 
will be maintained in a monoatomic chain. In fact, for a 
monoatomic linear chain with one electron per atom, the dimerized
state is more stable with a Peierls gap at the zone edge. The 
situation is expected to be more complex for the chain of aluminum atoms
having $3s^{2}3p^1$ valency.

The interest in metal nanowires is heightened by the observation
of quantized behavior of electrical conductance at room temperature
through connective necks stretching between two 
electrodes.\cite{agrait,pascu1,pascu2,olesen,krans1}
Studies attempting to simulate the process of stretching by using 
classical molecular dynamics have shown novel atomic and mechanical 
properties.\cite{pascu2,landm1,brotko,mehrez,sorens,finbow,gulser,tosatt} 
In particular, it was found\cite{mehrez} that the 2D hexagonal 
or square lattice structure of atomic planes perpendicular to 
the axis changes to the pentagons and later to equilateral 
triangles when the wire is thinned 
down to the radius of 5-10 \AA. Upon further thinning, strands 
(or bundles of finite atomic chains),\cite{mehrez} and eventually a 
monoatomic chain forms at the narrowest section of the 
nanowire.\cite{mehrez,sorens,finbow} 
Recently, the stability of suspended gold chains and their atomic 
structures have been studied
extensively.\cite{porta1,porta2,torres,okamot,hakkin,hakki2,spring} 

The first-principles calculations by Portal 
{\it et al.}\cite{porta1} showed that infinite, as well as 
finite gold atomic chains between two gold electrodes favor
the planar zigzag geometry at a bond angle $\alpha =$131$^o$.
The homogenization of the charge with a depletion in the interatomic
region ruled out the formation of a directional chemical bond.
On the other hand, the first-principles calculations by 
H\"{a}kkinen {\it et al.}\cite{hakkin,hakki2} for a finite gold 
chain between two gold electrodes favored the dimerized structure. 
In contrast to the conclusion drawn by Portal {\it et al.,}\cite{porta1} 
H\"akkinen {\it et al.}\cite{hakki2} attributed the stability of the 
suspended gold chain to the directional local bonding
with $spd$ hybridization. Apparently, the stability of a finite 
chain depends on the strain and the atomic configuration
where the chain is connected to the electrodes. In a more recent
comparative study\cite{porta2} Au, Cu, Ca, K infinite chains were
found to form planar zigzag structures with equilateral triangular 
geometry; only Au chain has a second zigzag structure with a wide bond 
angle $\alpha$=131$^o$. Note that these atoms can be considered similar 
because 
of their $s$-type outermost valence orbitals. Aluminum with $3s$-, and
$3p$-valence orbitals is different from Au, Cu, Ca. K.
Therefore, Al is an important element for understanding the 
formation and stability of ultimate 1D atomic chains. 

This paper presents a systematic, first-principle analysis of the binding, 
atomic and electronic structure of very thin Al chains. The objective 
is to reveal periodic linear, planar and non-planar geometries forming 
stable structures. An emphasis is placed on the planar structures forming 
zigzag chains. Some of the Al chain structures are compared with the
corresponding structures of Au chains. It is found that by going 
from bulk Al to a chain structure the character of bonding changes
and acquires directionality. The higher the coordination of individual
atoms, the stronger is the binding energy. It is hoped that the present
analysis will contribute to the understanding of atomic structure and
related physical properties (such as electrical and thermal conductance,
elascticity {\it etc.}) of infinite and finite atomic chains. 

\section{method}

First-principle calculations were carried out within the density 
functional theory. Al and Au chains are treated within the supercell 
geometry. To minimize the interchain interaction the distance between the

chains is taken to be 20 \AA. The wave functions are expressed by plane 
waves with the cutoff energy, $|{\bf k}+{\bf G}|^2 \leq $275 eV. The 
Brillouin zone (BZ) integration is performed within Monkhorst-Pack 
scheme\cite{monkh} using (1$\times$1$\times$40) ${\bf k}$-points. 
The convergence with respect to the energy cutoff and number of
{\bf k}-points were  tested.
Ionic potentials are represented by ultra soft Vanderbilt type 
pseudopotentials\cite{vander} and results are obtained by Generalized 
Gradient Approximation\cite{perdew} for fully relaxed atomic structures. 
Preconditioned conjugate gradient method is used for wave function
optimization. Since ionic relaxations are carried out by the conjugate
gradient method, the optimized (fully relaxed) structures obtained 
in this study are stable structures. In certain cases the stability
of a structure is tested by calculating the total energy while the
atoms are displaced in special directions. Numerical calculations are 
performed by using VASP code.\cite{hafner}
The $z$-axis is taken along the chain axis, and $y$-axis ($x$-axis) 
is perpendicular to (in) the plane of zigzag structure.

\section{Results}

\subsection{Optimized structures and cohesive energies}

The variation the total energy $E_T$, of the atomic Al chain calculated 
for the fully relaxed linear,  planar (zigzag and ladder), and
non-planar (cross) structures is shown in Fig.1. The geometries of these 
structures and their relevant structural parameters are shown by insets. 
Since the total energies are given with respect to the energy of the free
Al atom, the cohesive energy $E_C=-E_T$. The zigzag geometry displays
two minima; one occurs at $s=$1.26 \AA ~and has cohesive energy 
$E_C=$2.65 eV/atom; other has shallow minimum and occurs at 
$s=$2.37 \AA ~with cohesive energy $E_C=$1.92 eV/atom. 
The high cohesive energy zigzag structure 
(specified as $T$) having the bond length $d=$2.51 \AA, and the bond 
angle $\alpha \sim$60$^o$ forms equilateral triangles. This geometry 
allows for four nearest neighbors, which is less than the six nearest
neighbors occurring in the Al(111) atomic plane and twelve nearest 
neighbors in the close packed bulk metal. 
The equilateral triangular geometry can also be viewed as if two 
parallel linear chains with an interchain distance of 2.17 \AA~ are 
displaced by $d/2$ along the chain axis ($z$-direction). This is 
reminiscent of the hollow site registry of 2D atomic planes which
usually increases the cohesive energy.
 
The low cohesive energy zigzag structure (specified as $W$) has 
$d=$2.53 \AA~and wide bond angle $\alpha \sim $139$^o$, and allows 
for only two nearest neighbors with bonds slightly larger than 
those of the $T$-structure. We also found that the cohesive 
energy decreases if an Al atom is displaced perpendicular
to the zigzag plane. Therefore, both zigzag structures are planar. 
The minimum energy of the linear structure ($\alpha =$ 180$^o$ and 
denoted as $L$) has relatively short bond length, $d=s=$2.41 \AA. It is
$\sim$0.5 eV above the minimum energy of the $W$-structure and has 
cohesive energy $E_C=$1.87 eV/atom.

Two linear chains can form a ladder structure which allows for three
nearest neighbors with $\alpha=90^o$ and
$E_C \sim$2.4 eV/atom intermediate to the $T$-, 
and $W$-structures. The cohesive energy is further increased
to 2.5 eV/atom when the separation between chains is sligtly increasesd.
This way two strands (specified as $S$-structure) form, which are held
in place by the uniaxial stress between two electrodes.\cite{ohnish,mehrez}

Non-planar cross structure (specified as $C$-structure) has four atoms
which form two perpendicular dumbbells (A and B) in the unit cell. 
The lengths of these dumbells are different ( A: 2.8 \AA~ and B: 4.15 \AA)
and the chain is made by the ABABA... sequence of these dumbbells.
Al atoms in A has five non-planar bonds, and those in B have four 
bonds of $\sim$2.8 \AA. The cohesive energy of this structure 
is calculated to be 3.04 eV/atom. Since the atoms of the 
A-dumbbells are bound to the nearest five atoms 
forming equilateral triangles in different planes, and those of the
B-dumbbels have four non-planar bonds, this cohesive energy is 
highest among the 1D structures described in Fig. 1.
The $C$-structure was revealed first in the extensive analysis of
G\"{u}lseren {\it et al.} by using empirical glue potential.\cite{gulser} 
The overall features of the stable $C$-structure determined by these 
calculations are confirmed here, but the structural parameters are 
more accurately determined by the present first principle calculations.

It is worth noting  that the Au atomic chain also forms two different 
zigzag structures similar to those of Al; but Cu, Ca and K do 
not.\cite{porta2} Our calculated values for $s$, $d$, $E_C$ are 
respectively, 1.36 \AA, 2.71 \AA, 2.23 eV/atom
for the $T$-structure; 2.33 \AA, 2.56 \AA, 1.90 eV/atom for the 
$W$-structure; 2.59 \AA, 2.59 \AA, 1.68 eV/atom for the $L$-structure
of Au chain. The nearest neighbor distance of the $T$-structure of the Au 
chain is reduced only $\sim$ 6 $\%$ from that of the bulk. Is this puzzling 
similarity of 1D atomic structures of Al and Au (despite their dissimilar 
valencies) only a coincidence? We now address this issue.

We calculated the cohesive energy of bulk Al (Au), $E_C=$3.67 (3.20)
eV/atom at the nearest neighbor distance $d=$2.86 (2.96) \AA~(or 
lattice parameter $a=$4.04 (4.18) \AA).\cite{table} The energetics of 
1D and bulk structures are compared in Table I. Simple arguments 
based on the counting of nearest neighbor couplings would suggest a 
relatively small cohesive energy, e.g., $\sim$ 1.3 eV for the 
$T$-structure. On the contrary, 1D structures studied here have cohesive 
energies higher than one can estimate by comparing their coordination 
numbers with that of bulk. Apparently, the bonds in 1D structures
become stronger. In fact, it was found previously that the 
linear Al chain has a Young's modulus stronger than bulk.\cite{mehrez} 
Recent scanning tunneling microscope studies revealed that the bond 
strength of the Au nanowire is about twice that of a bulk metallic 
bond.\cite{bollig}

\subsection{Charge density analysis}

Figure 2 shows the charge density contour plots of bulk, $L$-, $W$-, and 
$T$-structures. In contrast to uniform metallic charge density of bulk,
the bonding acquires directionality in 1D structures of Al. For the 
$L$-structure
the charge is accumulated between atoms forming a directional bond, and 
is mainly due to the $\sigma$ states (formed by $3s+3p_z$ orbitals)
and partly due to $\pi$ states (formed by $3p_x$ and
$3p_y$ orbitals perpendicular to the chain axis). The calculated charge 
distribution suggests that directional ``covalent'' bonds are responsible
for the bonding. This situation is maintained in the zigzag $W$-structure
except for a slight distortion of the bond charge. Actually, the 
$W$-structure with wide bond angle is not dramatically different
from the $L$-structure. In the $T$-structure that forms
equilateral triangles, the charge density is apparently different from
that of the $W$-structure. We see a continuous (connected) region
of high charge density between double atomic chains. However, this
is nothing but the overlap of charges of four bonds emerging from
each chain atom, and is confirmed by the contour plot of an 
individual bond charge in a plane perpendicular to the zigzag ($xz$) 
plane and passing through an Al--Al bond.
We also notice that the charge becomes slightly delocalized by
going from $L$ to $T$-structure. These charge distributions of the 
Al--Al bond described above is different from 
the corresponding charge distribution of Au zigzag 
structures shown in Fig. 3. Clearly, there are no directional bonds 
in the Au chain; valence charge is delocalized. This finding is in
confirmity with the results of Portal {\it et al.}\cite{porta1,porta2}
on the infinite Au chain. On the other hand, 
H\"{a}kkinen {\it et al.}\cite{hakki2} deduced directional
bonding with $spd$ hybridization in finite Au chains between two Au 
electrodes by performing similar type of pseudopotential plane wave 
calculations. 

\subsection{Electronic structure}

A comparative analysis of the electronic band structure of Al monoatomic 
chains illustrated in Fig. 4 provides further insight into the stability
and character of bonding. The band structure of the $L$-structure is 
folded for the sake of comparison with the zigzag structures. Two filled 
$\sigma$ bands arise from the $3s+3p_z$ valence orbitals and make
the bond charge shown in Fig. 2a. Because of the linear geometry
$3p_x$ and $3p_y$ are equivalent, and give rise to doubly degenerate
$\pi$ band crossing the Fermi level. As pointed out by Peierls,\cite{peierl}
a one-dimensional metal with a partly filled band will distort away from
a regular chain structure to lower its energy. According to the above 
analysis,
a linear chain of uniformly spaced Al atoms with spacing $s=d$ has a 
quarter-filled band which crosses the Fermi level at $k_{z}=\pm \pi/4 d$.
A distorted unit cell $4d$ in length will cause this point to coincide
with the edge of the Fermi distribution. The change in the crystal
potential due to the $4d$ distortion will open up a Peierls gap at the
reduced zone-edge lowering the total energy. In practice, the gain in 
energy due to such Peierls distortions is rather small even for a $2d$
dostortion (dimerization) and is likely to be below computational 
error for the $4d$ distortion here. Thus, although the linear chain of Al
atoms is unstable, in principle, the effect of such a distortion on 
cohesive energy is clearly negligible.

The symmetry between $3p_x$ and
$3p_y$ orbitals is broken in the zigzag structure, and hence the
$\pi$-band is split. Apart from this band splitting and slight 
rise of bands, the overall form of the energy band structure is 
maintained in the $W$-structure. The $W$-structure is, however, more
stable than the $L$-structure because of its relatively stronger
electronic screening. In the $T$-structure the split $\pi$-bands are 
lowered, and the form of the $\sigma$-bands undergo a significant 
change due to the equilateral triangular geometry. Despite slight 
delocalization of charge, the total energy of the 
$T$-structure is lower than the $W$-structure. The relative
stability originates from the increased number of nearest neighbors.
It is noted that the bands in all chain structures can be considered
similar as far as Fermi level crossing of the bands is concerned.

In contrast to the Al chains described above, the energy band structure 
of Au undergoes significant changes in different 
structures near the Fermi energy.
For example, for the linear structure one band crosses the Fermi level
near the $X$-point of the BZ, another band at the $\Gamma$-point is very 
close to the Fermi level. For the $W$-structure, two bands cross at the 
Fermi level and at the zone boundary with negligible Peierls 
distortion gap\cite{batra1}
and the rest of the bands are lowered. The lowering of the state density
at the Fermi level stabilizes the zigzag structure relative to the
linear structure.\cite{porta1,porta2} In the $T$-structure two bands cross
the Fermi level. Inspite of these changes the character of the bonding
remains essentially metallic in the chain structures.

\section{Discussion and conclusions}

The analysis of the above results and comparison with Au chain reveal that
the metallic bond of bulk Al changes to a directional covalent bond in the
1D monoatomic chain. The metallicity is ensured by the $\pi$-states. 
For that reason, our efforts of calculating the bands
of the Al chains with tight binding method by using the bulk 
parameters\cite{papaco} have not been too successful. This suggests 
that the transferability of energy parameters fitted to bulk is not 
satisfactory for the 1D ($T$-, $W$- and $L$-) structures. 

The $T$ structure with two parallel linear chains can be 
viewd as the 1D analog of the 3D close packing. In this 
respect, the $T$-structure may be considered in a different 
class and as a precursor of the 2D hexagonal lattice. 
Adding one more parallel chain in registry with the quasi 1D 
$T$-structure, one starts to build the hexagons, where 2/3 of the atoms 
have four and 1/3 of the atoms have six nearest neighbors. 
As a natural extention of these arguments, another intermediary,
quasi 1D structure, for example, is a ladder structure which consists of
two parallel linear chains forming a row of squares with a lattice 
constant of $d$ and allowing for three nearest neighbors. 
This metastable structure is a 1D analog of the top site registry 
of 2D atomic planes.\cite{cirac5} Our calculations 
show that the cohesive energy of the ladder structure is 
increased when the distance between two chains increases, so that the
chain turns to two strands. The cohesive energy of the strands is found 
between the $T$- and $W$-structure. By going from planar to non-planar
geometry the cohesive energy further increases
The present work suggests that the 
Born-Oppenheimer surface for these quasi 1D structures is rather complex, 
and generally $E_C$ increases with increasing coordination number 
and decreasing bond angle. As clarified in Sec. II, the stable structures
correspond to the local minima on the Born-Oppenheimer surface and are
expected to be vibrationally stable at least at low temperatures. The 
1D $T$-structure found for Al, Au, Cu, Ca and K appears to be common to
metals, in a way an intermediate structure between a truly 1D and
2D structures.

We also note that the linear and zigzag structures of Al have
two bands crossing the Fermi energy. Calculations using the Green's
function method\cite{alper0} yield one conduction channel for each band of 
uniform chain crossing the Fermi level, and hence the ballistic conductance
of the $T$-, $W$-, and $L$-structures $G=2(2e^2/h)$. This value for the 
conductance arises from the fact that the channel capacity or the 
maximum conductance per channel is $2e^2/h$. It is straightforward to 
motivate the maximum value by appealing to the Heisenberg's uncertainity 
principle.\cite{batra2} Recalling that
conductance $G=\Delta I/ \Delta V$, and $\Delta I=\Delta Q/\Delta t$, then
for a single channel in extreme quantum limit $\Delta Q=e$. One can readily
write $G=e^2/\Delta E \Delta t$. Now invoking the uncertainity principle,
$\Delta E \Delta t \geq h$, one finally obtains $G \leq 2e^2/h$. Here the 
factor of two is due to spin. The maximum conductance per channel can
never be greater than $2e^2/h$. We note that the value of the ballistic 
conductance for the infinite chain ({\it i.e.} $4e^2/h$) is at 
variance with the experimental results\cite{scheer} yielding only
$G \sim 2e^2/h$ for the finite Al chain. The discrepancy was explained
by the fact that the electronic states are modified due to the 
finite size of the 
chain and the atomic configuration where the chain is coupled to the 
electrodes. \cite{mehre2} 

It is important to remark that creation of metallic overlayers on
semiconductors is required in chip technology. Thus if one could
place these quasi 1D structures on semiconductor surfaces without
losing the metallic behavior of the chains, the technology would
be considerably enhanced. Unfortunately, when one examines the 
deposition of monolayers of metals like Al, Au and Ga on Si, the 
lowest energy configuration turns out to be semiconducting in 
nature. A metastable state, in which Al forms a metallic zigzag 
structure on the Si(100) surface, has been reported.\cite{batra3} 
It remains to be seen how feasible it is to fabricate such a 
structure.

In summary, we have found that a zigzag chain of aluminum in triangular 
configuration is most stable among the planar structures we studied. 
The structural results for planar geometries are similar to gold
but bonding is different. A new metastable ladder structure intermediate
to distorted linear and triangular structure is also reported. The 
metallicity
has its origin in the $\pi$-bands. The stabilization of these metallic
monoatomic chains on semiconductors remain an experimental challenge.

\section{acknowledgement} We thank Dr Oguz G\"{u}lseren for his 
stimulating discussions.

\begin{figure}
\caption{The calculated total energy $E_T$ (and also cohesive energy,
$E_C$) of an infinite Al chain with linear, planar (zigzag and ladder),
and non-planar (cross) strucures. 
Energies are given relative to the energy of a free Al atom. 
The calculated energy of bulk Al is indicated by arrow. Relevant 
structural parameters, bond length $d$, bond angle $\alpha$,
$s=$ and $h$ are shown by inset for non-planar cross $C$,
high energy (or equilateral 
triangular) $T$, low energy zigzag $W$, ladder (or strands) $S$
and linear ($L$) geometries. The zigzag structure is in the ($xz$)-plane.
The short and long dumbbells of the $C$-structure are along $x$- and 
$y$-axis, respectively. For values of energies and structural
parameters see Table I.} 
\end{figure}

\begin{figure}
\caption{Charge density counterplots of 1D and 3D Al structures: (a)
Bulk; (b) Linear geometry; (c) $W$-geometry, on the plane of the 
zigzag structure ({\it i.e.} $xz$-plane); (d) $W$-geometry, on the plane 
passing through the Al--Al bond and perpendicular to the plane
of the zigzag structure; (e) $T$-geometry, on the plane of the zigzag 
structure; (f) $T$-geometry, on the plane passing through the bond and 
perpendicular to the plane of the zigzag structure.
Increasing direction of the charge density is indicated by arrows. Numerals
show the highest contour values. Atomic positions are indicated by {\bf x}.}
\end{figure}

\begin{figure}
\caption{Charge density counterplots of the Au chain. (a)
$T$-structure; (b) $T$-structure, on the
plane passing through the bond and perpendicular to the zigzag plane.}
\end{figure}

\begin{figure}
\caption{Energy band structure of Al and Au chains. (a) Linear $L$-;
(b) low energy zigzag, $W$-; (c) high energy zigzag $T$-structure
of Al. (d) Linear $L$-; (e) $W$-; (f) $T$-structure of Au. Bands of 
$L$-structure is zone folded for the sake of comparison
with the zigzag structures. Zero of energy is taken at the Fermi level.}
\end{figure}

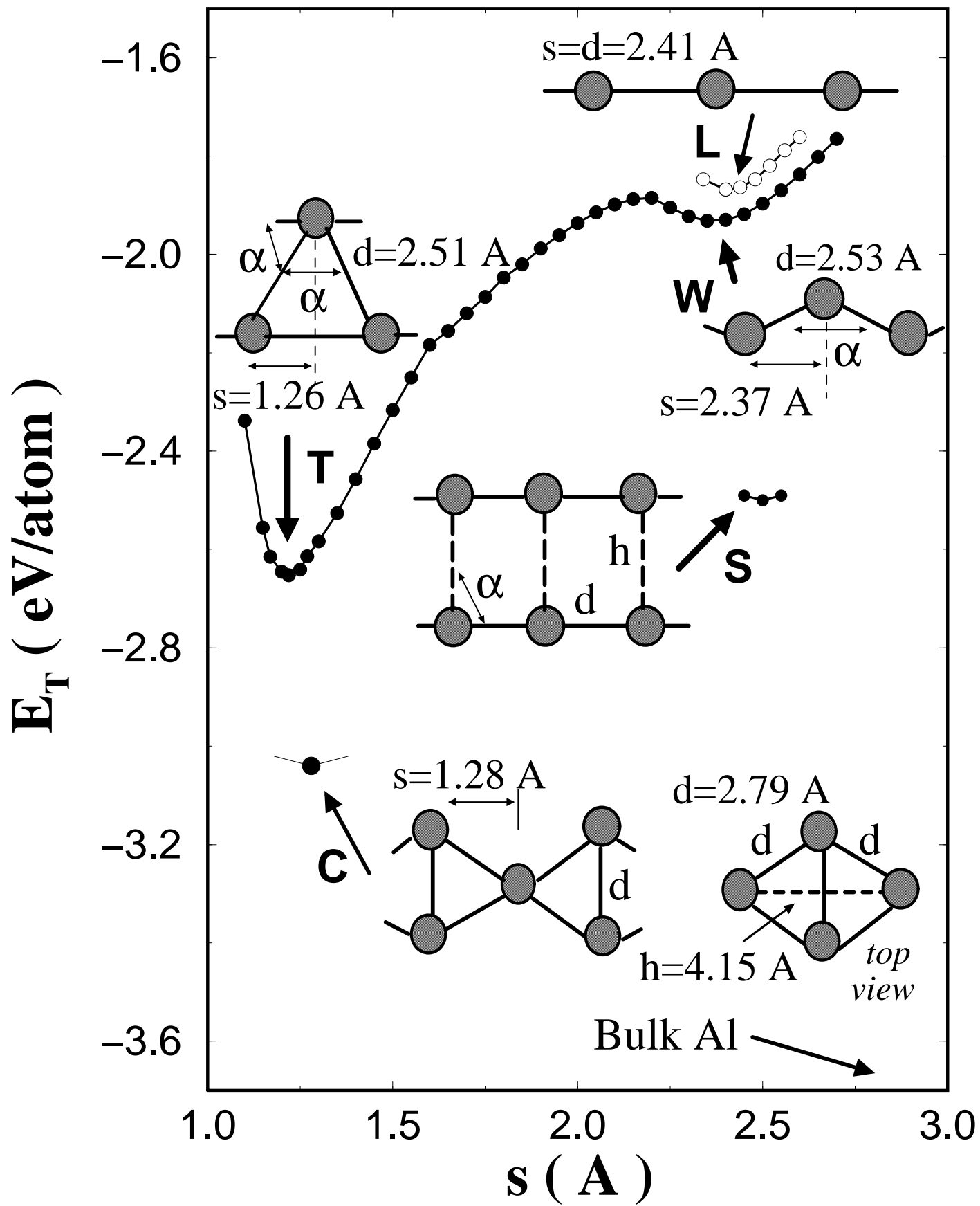
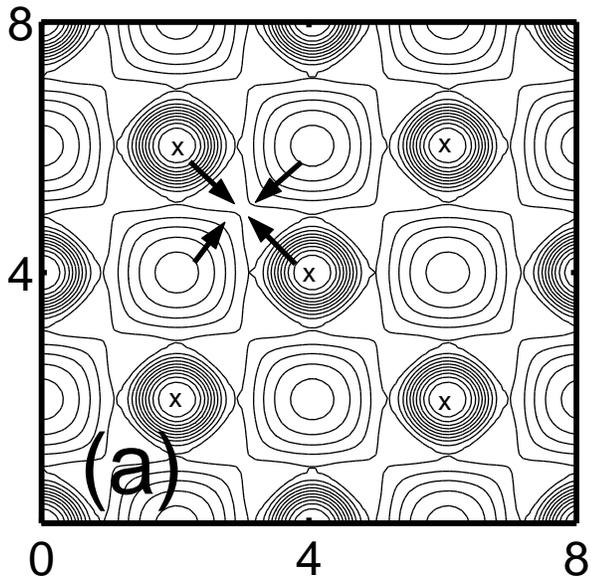
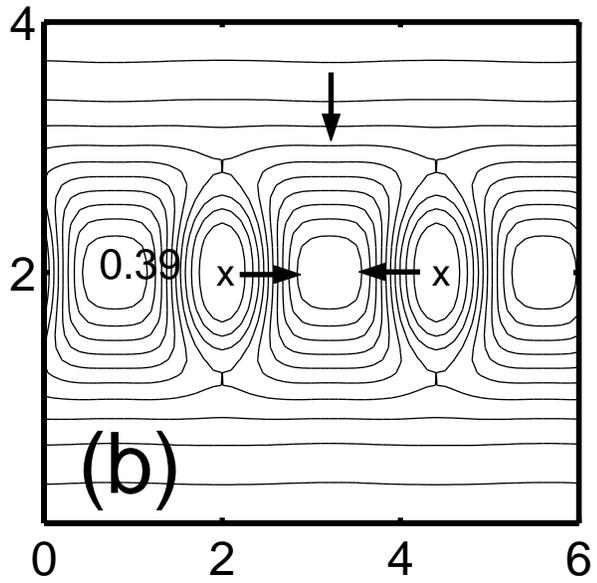
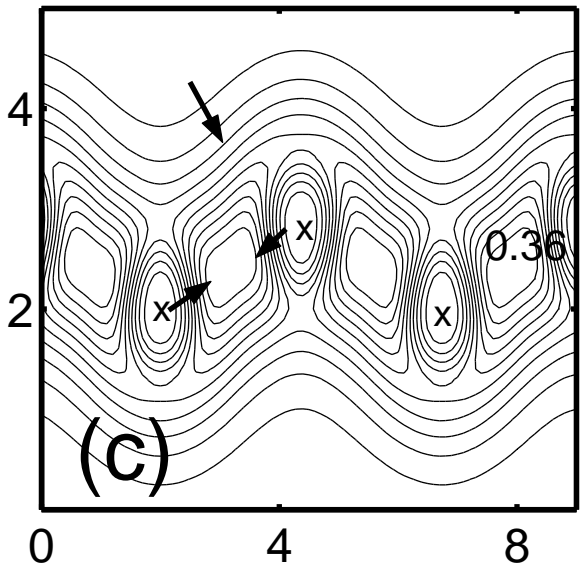
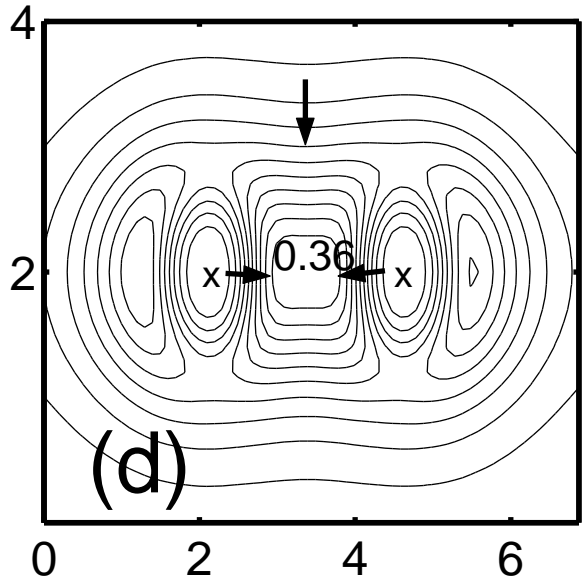
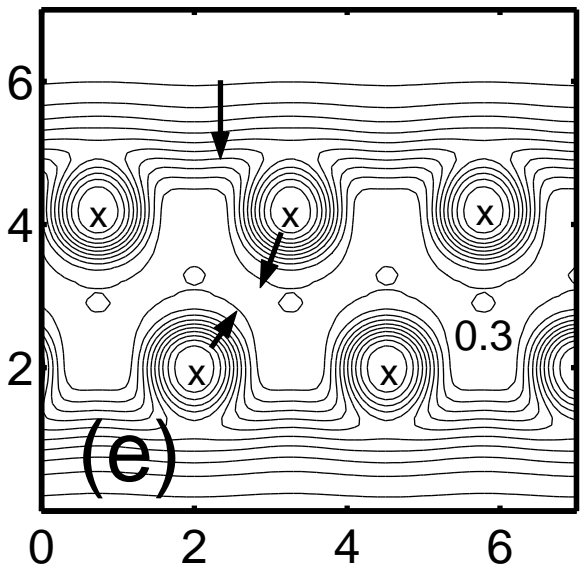
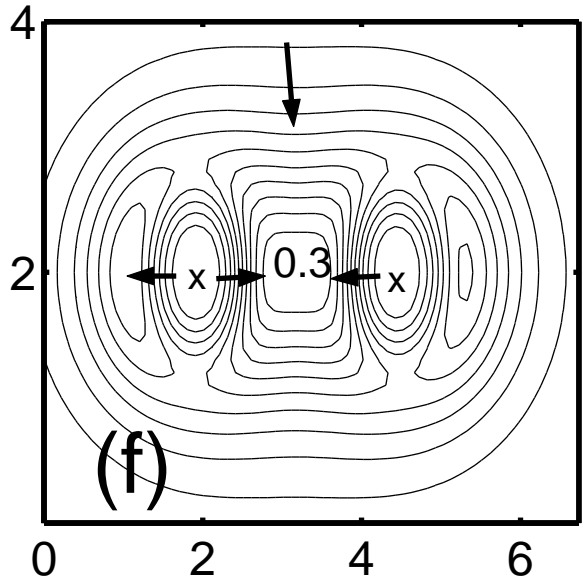
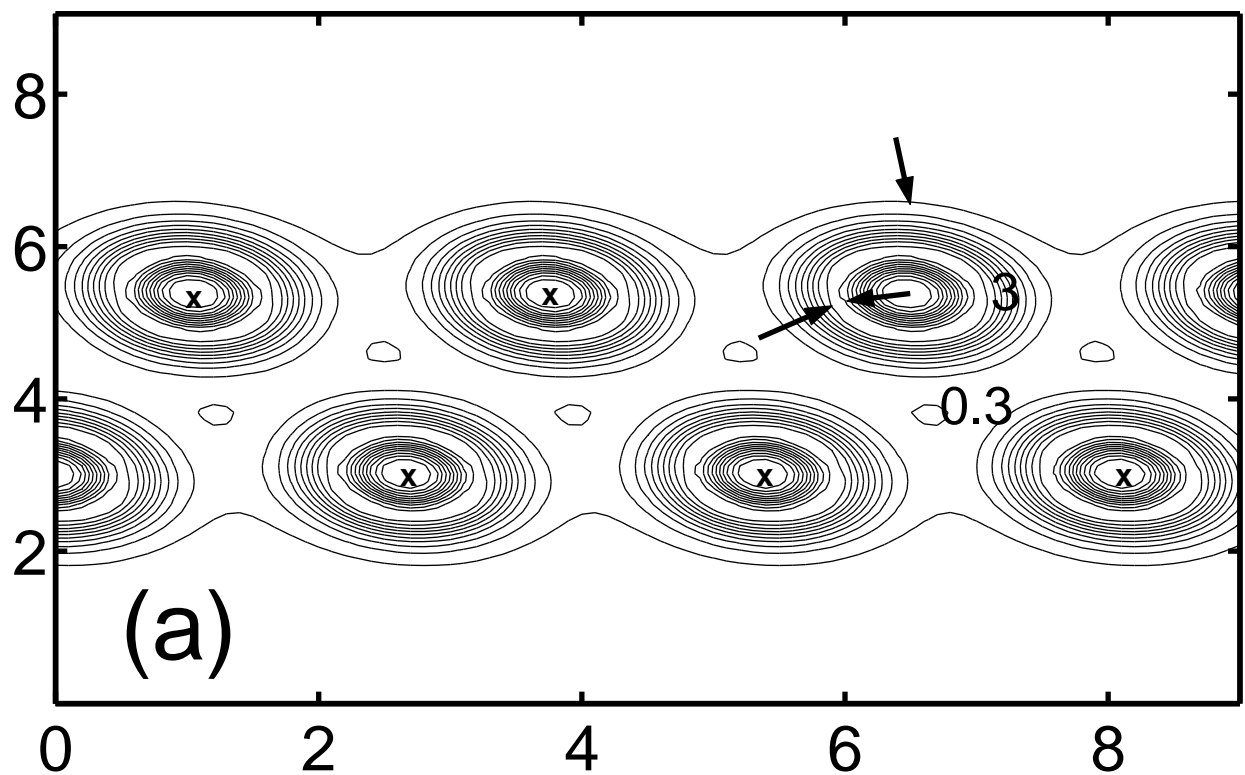
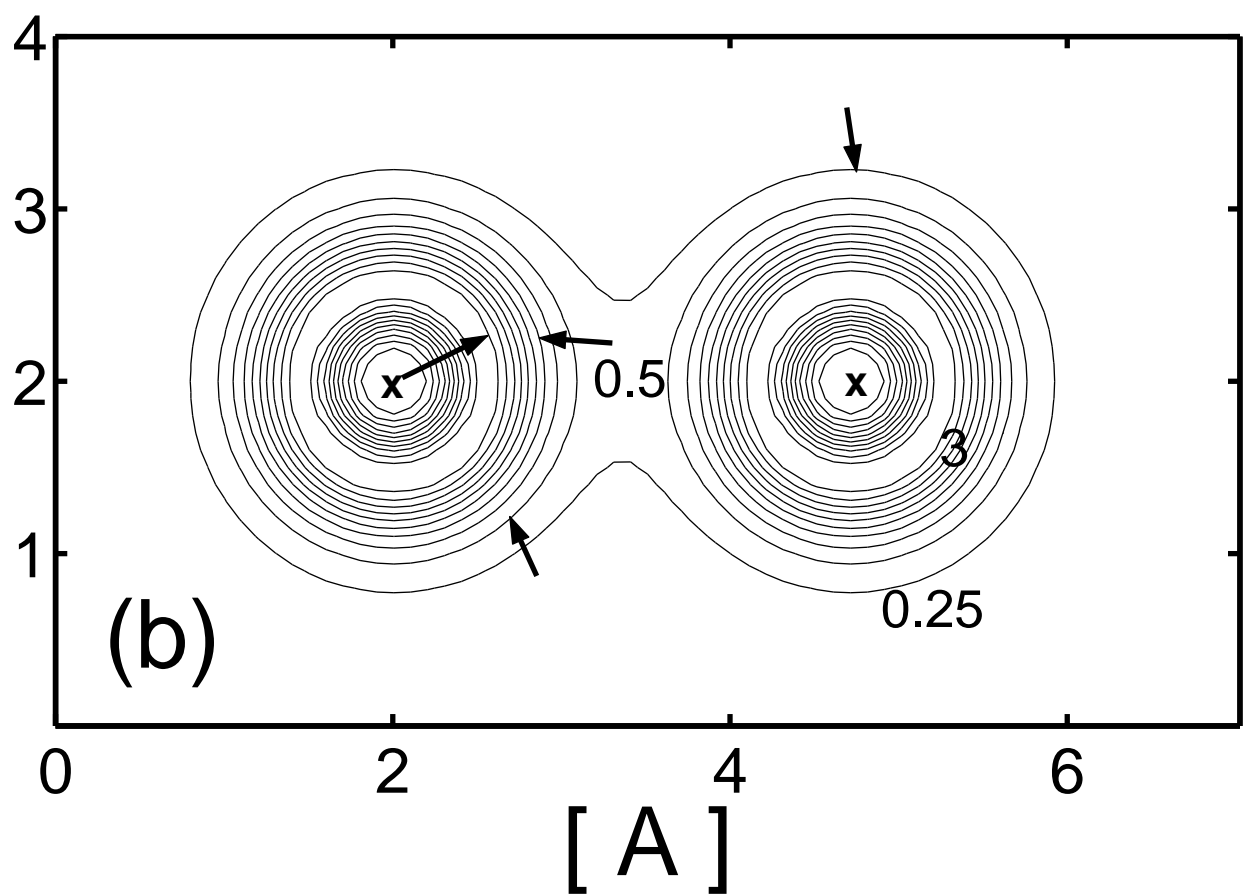
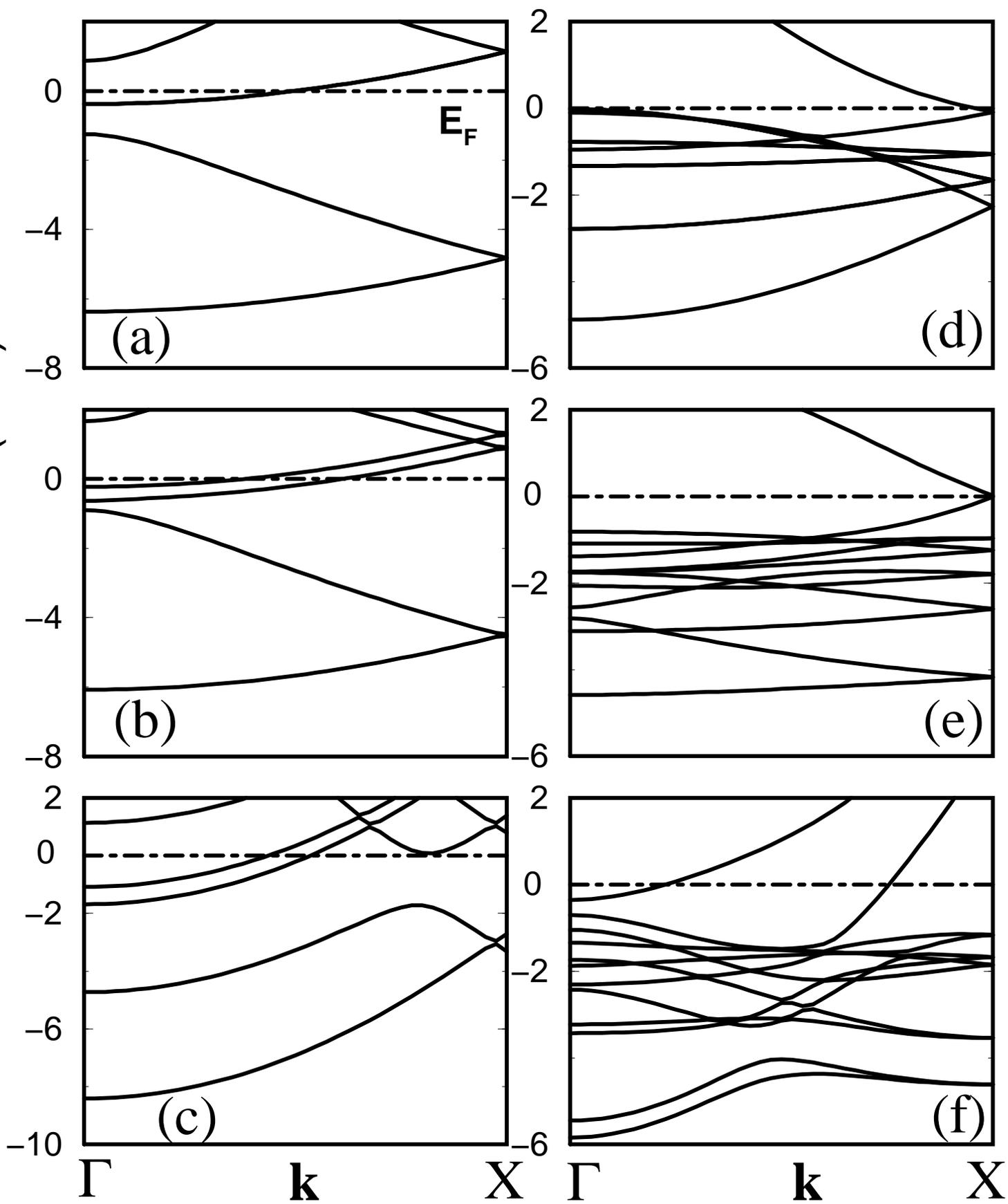
\end{document}